\documentclass[twocolumn,showpacs,aps,prl,superscriptaddress,floatfix]{revtex4}

\usepackage{graphicx}
\usepackage{dcolumn}
\usepackage{amsmath}
\usepackage{epsfig}
\usepackage{multirow}

\newcommand{\BaBarYear}       {04}
\newcommand{\BaBarNumber}     {19}
\newcommand{\SLACPubNumber} {10529}

\def\de   {\Delta E^*}

\def\mes  {M_{ES}}

\def\bkg  {B \to K^{*}\gamma}

\def\bkgneut  {B^0 \to K^{*0}\gamma}
\def\bkpg  {\bu \to K^{*+}\gamma}

\def\bpdpi {B^- \to D^0\pim}
\def\bzdpi {B^0 \to D^-\pip}

\def\bsg  {b\to s\gamma}

\def\kstar {K^*}

\def\incbsg  {B\to X_s\gamma}

\def\acp {\ensuremath{\mathcal{A}}}

\def\bkg    {\ensuremath {\B \to \Kstar \gamma}}
\def\bkog    {\ensuremath {\Bz \to \Kstarz \gamma}}

\def\bkpg    {\ensuremath {\Bp \to \Kstarp \gamma}}

\def\Kz    {\ensuremath{K^{0}}\xspace}

\def\de        {\ensuremath {\Delta E^{*}}}

\def\MC        {MC}

\RequirePackage{xspace}

\usepackage{relsize}
\def\babar{\mbox{\slshape B\kern-0.1em{\smaller A}\kern-0.1em
    B\kern-0.1em{\smaller A\kern-0.2em R}}}

\def\qqbar {\ensuremath{q\overline q}\xspace}

\def\piz   {\ensuremath{\pi^0}\xspace}

\def\pip   {\ensuremath{\pi^+}\xspace}
\def\pim   {\ensuremath{\pi^-}\xspace}

\def\Kbar  {\kern 0.2em\overline{\kern -0.2em K}{}\xspace}

\def\Kz    {\ensuremath{K^0}\xspace}
\def\Kzb   {\ensuremath{\Kbar^0}\xspace}
\def\KzKzb {\ensuremath{\Kz \kern -0.16em \Kzb}\xspace}
\def\Kp    {\ensuremath{K^+}\xspace}
\def\Km    {\ensuremath{K^-}\xspace}

\def\KpKm  {\ensuremath{\Kp \kern -0.16em \Km}\xspace}
\def\KS    {\ensuremath{K^0_{\scriptscriptstyle S}}\xspace} 
 
\def\Kstarz  {\ensuremath{K^{*0}}\xspace}

\def\Kstar   {\ensuremath{K^*}\xspace}

\def\Kstarp  {\ensuremath{K^{*+}}\xspace}

\def\Dbar    {\kern 0.2em\overline{\kern -0.2em D}{}\xspace}

\def\Dz      {\ensuremath{D^0}\xspace}
\def\Dzb     {\ensuremath{\Dbar^0}\xspace}
\def\DzDzb   {\ensuremath{\Dz {\kern -0.16em \Dzb}}\xspace}
\def\Dp      {\ensuremath{D^+}\xspace}
\def\Dm      {\ensuremath{D^-}\xspace}

\def\DpDm    {\ensuremath{\Dp {\kern -0.16em \Dm}}\xspace}

\def\B       {\ensuremath{B}\xspace}
\def\Bbar    {\kern 0.18em\overline{\kern -0.18em B}{}\xspace}

\def\BB      {\ensuremath{B\Bbar}\xspace} 
\def\Bz      {\ensuremath{B^0}\xspace}
\def\Bzb     {\ensuremath{\Bbar^0}\xspace}
\def\BzBzb   {\ensuremath{\Bz {\kern -0.16em \Bzb}}\xspace}
\def\Bu      {\ensuremath{B^+}\xspace}
\def\Bub     {\ensuremath{B^-}\xspace}
\def\Bp      {\ensuremath{\Bu}\xspace}

\def\BpBm    {\ensuremath{\Bu {\kern -0.16em \Bub}}\xspace}

\def\BorBbar    {\kern 0.18em\optbar{\kern -0.18em B}{}\xspace}
\def\DorDbar    {\kern 0.18em\optbar{\kern -0.18em D}{}\xspace}
\def\KorKbar    {\kern 0.18em\optbar{\kern -0.18em K}{}\xspace}

\mathchardef\Upsilon="7107
\def\Y#1S{\ensuremath{\Upsilon{(#1S)}}\xspace}

\mathchardef\Deltares="7101
\mathchardef\Xi="7104
\mathchardef\Lambda="7103
\mathchardef\Sigma="7106
\mathchardef\Omega="710A

\def\Deltabar{\kern 0.25em\overline{\kern -0.25em \Deltares}{}\xspace}
\def\Lbar{\kern 0.2em\overline{\kern -0.2em\Lambda\kern 0.05em}\kern-0.05em{}\xspace}
\def\Sigbar{\kern 0.2em\overline{\kern -0.2em \Sigma}{}\xspace}
\def\Xibar{\kern 0.2em\overline{\kern -0.2em \Xi}{}\xspace}
\def\Obar{\kern 0.2em\overline{\kern -0.2em \Omega}{}\xspace}
\def\Nbar{\kern 0.2em\overline{\kern -0.2em N}{}\xspace}
\def\Xb{\kern 0.2em\overline{\kern -0.2em X}{}\xspace}

\def\BR         {{\ensuremath{\cal B}\xspace}}

\def\mes        {\mbox{$m_{\rm ES}$}\xspace}

\def\DeltaE     {\mbox{$\Delta E$}\xspace}

\newcommand{\tev}{\ensuremath{\mathrm{\,Te\kern -0.1em V}}\xspace}
\newcommand{\gev}{\ensuremath{\mathrm{\,Ge\kern -0.1em V}}\xspace}
\newcommand{\mev}{\ensuremath{\mathrm{\,Me\kern -0.1em V}}\xspace}
\newcommand{\kev}{\ensuremath{\mathrm{\,ke\kern -0.1em V}}\xspace}
\newcommand{\ev}{\ensuremath{\mathrm{\,e\kern -0.1em V}}\xspace}
\newcommand{\gevc}{\ensuremath{{\mathrm{\,Ge\kern -0.1em V\!/}c}}\xspace}
\newcommand{\mevc}{\ensuremath{{\mathrm{\,Me\kern -0.1em V\!/}c}}\xspace}
\newcommand{\gevcc}{\ensuremath{{\mathrm{\,Ge\kern -0.1em V\!/}c^2}}\xspace}
\newcommand{\mevcc}{\ensuremath{{\mathrm{\,Me\kern -0.1em V\!/}c^2}}\xspace}

\def\mus  {\ensuremath{\rm \,\mus}\xspace}

\def\mus        {\ensuremath{\,\mu{\rm s}}\xspace}    

\def\to                 {\ensuremath{\rightarrow}\xspace}

\def\pep2{PEP-II}

\def\gsim{{~\raise.15em\hbox{$>$}\kern-.85em
          \lower.35em\hbox{$\sim$}~}\xspace}
\def\lsim{{~\raise.15em\hbox{$<$}\kern-.85em
          \lower.35em\hbox{$\sim$}~}\xspace}

\def\CP                {\ensuremath{C\!P}\xspace}

\xspace

\def\jetset74   {\mbox{\tt Jetset \hspace{-0.5em}7.\hspace{-0.2em}4}\xspace}

\begin{document}

{\pagestyle{empty}
\begin{flushleft}
\babar-PUB-\BaBarYear/\BaBarNumber \\
SLAC-PUB-\SLACPubNumber \\
\end{flushleft}
}

\title{Measurement of Branching Fractions, and CP and Isospin Asymmetries, for $B \to K^* \gamma$}

\def\brcharg {\ensuremath {\BR(\bkpg) = (3.87 \pm 0.28({\rm stat. }) \pm 0.26({ \rm syst. })) \times 10^{-5}}}
\def\brneut {\ensuremath {\BR(\bkog)= (3.92 \pm 0.20({\rm stat. }) \pm 0.24({ \rm syst. })) \times 10^{-5}}}
\def\brsm {\ensuremath {\BR(\bkg)= 4.1 \times 10^{-5}}}
\def\aall {\ensuremath {\acp(\bkg) = -0.013 \pm 0.036({\rm stat. }) \pm 0.010({ \rm syst. }) }}
\def\aexcl {\ensuremath {  -0.074 <\acp(\bkg) <  0.049 }}
\def\isoval{\ensuremath {\Delta_{0-} = 0.050\pm 0.045({\rm stat. }) \pm 0.028({ \rm syst.}) \pm 0.024 (R^{+/0}) }}
\def\isoexcl{\ensuremath {-0.046 < \Delta_{0-} < 0.146 }}

\begin{abstract}
\noindent
The branching fractions of the decays $\bkog$ and $\bkpg$ are measured
using a sample of $\mbox{88}\times \mbox{10}^6 B\overline{B}\ $ events
collected with the \babar\ detector at the \pep2\ asymmetric-energy
$e^{+}e^{-}$ collider.  We find $\brneut$, $\brcharg$.  Our
measurements also constrain the direct $\CP$ asymmetry to be $\aexcl$
and the isospin asymmetry to be $\isoexcl$, both at the 90\%
confidence level.
\end{abstract}

\pacs{13.25.Hw,12.15.Hh,11.30.Er}

%
\author{B.~Aubert}
\author{R.~Barate}
\author{D.~Boutigny}
\author{F.~Couderc}
\author{J.-M.~Gaillard}
\author{A.~Hicheur}
\author{Y.~Karyotakis}
\author{J.~P.~Lees}
\author{V.~Tisserand}
\author{A.~Zghiche}
\affiliation{Laboratoire de Physique des Particules, F-74941 Annecy-le-Vieux, France }
% ...
\collaboration{The \babar\ Collaboration}
\noaffiliation

\maketitle



Within the Standard Model (SM), the decays \mbox{\bkg} proceed
dominantly through one-loop \mbox{$\bsg$} electromagnetic ``penguin''
transitions \cite{penguin:review}.  Non-SM virtual particles may be
present in these loops, changing the decay rates from the SM
predictions.  Theoretical calculations of exclusive \mbox{\bkg} decay
rates have large uncertainties due to nonperturbative hadronic effects
~\cite{Beneke:2001at,Bosch:2001gv,Ali:2001ez}, limiting their
usefulness for probing new physics.  Previous measurements
\cite{Coan:1999kh,Aubert:2001me,Nakao:2004} of the branching fractions
are already more precise than SM-based theoretical estimates, and are
in reasonable agreement with them.  Calculations
~\cite{DelDebbio:1998kr,Ball:1998kk} of the form factor for $\bkg$ can
be tested using improved measurements of these branching fractions.

Much of the theoretical uncertainty in the branching fractions cancels
in the ratios defining the isospin asymmetry $\Delta_{0-}$ and the CP
asymmetry $\acp$:

\begin{equation}
\Delta_{0-} =
{
{\Gamma(\overline{B}^0 \to \overline{K}^{*0}\gamma) - \Gamma(B^{-}\to K^{*-}\gamma)} \over
{\Gamma(\overline{B}^0 \to \overline{K}^{*0}\gamma) + \Gamma(B^{-}\to K^{*-}\gamma)}
},
\label{eq:isospin}
\end{equation}

\begin{equation}
\acp = 
{
{\Gamma(\overline{B} \to \overline{K}^*\gamma) - \Gamma(\bkg)} \over
{\Gamma(\overline{B} \to \overline{K}^*\gamma) + \Gamma(\bkg)}
},
\label{eq:acp}
\end{equation}

\noindent
making them stringent tests of the SM.  A further advantage of these
asymmetries is that some experimental systematic uncertainties cancel
in the ratios.  The SM predicts a positive value of $\Delta_{0-}$
between $5$ and $10\%$~\cite{Kagan:2001zk}, and $|\acp|$ less than
$1\%$~\cite{Kagan:1998bh}.  New physics contributions can modify these
values significantly ~\cite{Kagan:2001zk,Kagan:1998bh}.

In this Letter, we present measurements of the exclusive branching
fractions $\BR(\bkog)$ and $\BR(\bkpg)$, the isospin asymmetry
($\Delta_{0-}$), and the CP asymmetries $\acp(\bkog)$ and
$\acp(\bkpg)$.  $\Kstar$ refers to the $\Kstar(892)$ resonance
throughout this paper.  Inclusion of charge-conjugate decays is
implied except in the definitions of $\acp$.  This analysis uses
$(\mbox{88} \pm 1)\times \mbox{10}^6 B\overline{B}\ $ events, from
$\Upsilon(4S)$ decays, recorded by the $\babar$
detector~\cite{Aubert:2001tu}.  An additional 10 fb$^{-1}$ of data,
taken 40 MeV below the $\Upsilon(4S)$ resonance, is used for studying
non-$B$ continuum background.  After $\bkg$ event reconstruction and
background rejection, multi-dimensional extended maximum likelihood
fits are used to extract the final results.

We reconstruct $\bkog$ in the \mbox{$\Kstarz \to \Kp \pim$},
\mbox{$\KS\piz$} modes and $\bkpg$ in the \mbox{$\Kstarp \to \Kp
\piz$}, \mbox{$\KS \pip$} modes as described in detail in
Ref. ~\cite{Aubert:2001me, Aubert:2001tu}.  Reconstructed tracks are
identified as final state $\pi^\pm$ and $K^\pm$ mesons by measuring
the angle of the Cherenkov cone and energy loss along the track
(dE/dx).  The $\KS$ candidates are composed from pairs of oppositely
charged tracks with an invariant mass that is within
$\mbox{3.3}\sigma$ of the nominal $\KS$ mass and with a vertex that is
at least 0.3 cm away from the primary event vertex.  The
$\pi^0$-candidate momentum vector is determined by a mass-constrained
fit to pairs of photons, reconstructed from energy deposits in the
calorimeter that are not matched to tracks.  The $K$ and $\pi$
candidates are combined to form $\Kstar$ candidates, which are
required to have invariant mass in the range $800 < M_{K \pi} < 1000
\mevcc$.  The primary-photon candidates are required to have high
center-of-mass (CM) energy, between 1.5 and 3.5 \gev, and to satisfy
additional requirements designed to suppress the large $\piz$ and
$\eta$ background as described in Ref. \cite{Aubert:2001me}.

The $B$-meson candidates are reconstructed by combining the \Kstar\
and high-energy photon candidates.  We define in the CM frame (denoted
by asterisks) $\de \equiv E^*_{B}-E_{\rm beam}^*$, where $E_{\rm
beam}^*$ is the beam energy, known to high precision, and
$E^*_B=E^*_{\gamma}+E^*_{K^*}$ is the energy of the $B$-meson
candidate.  We also define the beam-energy-substituted mass $\mes
\equiv \sqrt{ E^{*2}_{\rm beam}-\mbox{$\boldmath{\mathrm
p'}$}_{B}^{*2}}$, where ${\mathrm p'}_B^*$ is the momentum of the $B$
candidate modified by scaling the photon energy to make
$E^*_\gamma+E^*_{K^*}-E_{beam}^*=0$.  This procedure reduces the tail
in the signal $\mes$ distribution, which results from the asymmetric
calorimeter response.  For signal decays, this ``rescaled'' $\mes$
peaks near 5.279$\gevcc$ with a resolution of $\approx$3 \mevcc\ and
$\de$ peaks near 0$\mev$ with a resolution of $\approx$50 \mev. We
consider only candidates with $\mes\ >\ \mbox{5.20}\gevcc\ $ and
$|\de|\ <\ \mbox{0.3 \gev}$.

\begin{figure*}[tb]
\includegraphics[width=\linewidth]{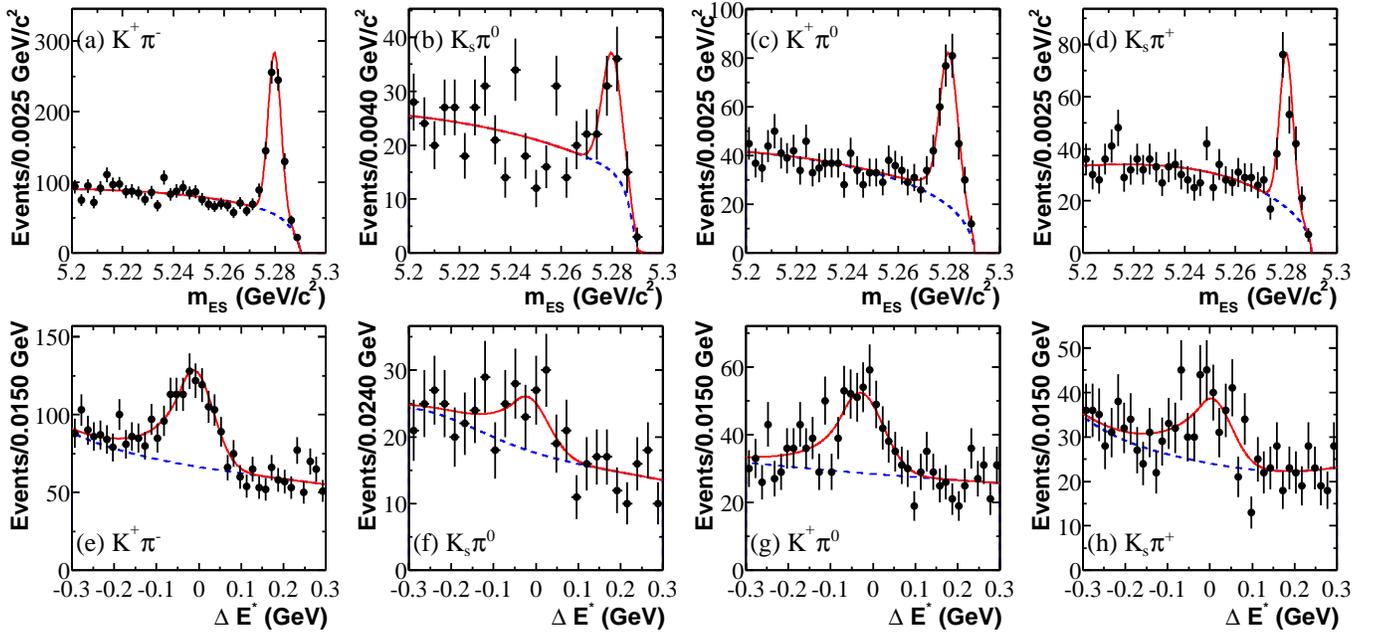}
\caption{\mes\ and \de distributions for the \bkg\ candidates. The
points are data, and the solid and dashed curves show the projections
of the complete fit and the background component alone,
respectively. The fits used to extract the signal yields are described
in the text.}
\label{fig:mes}
\end{figure*}

\begin{figure}[hbtp]
\noindent
\includegraphics[width=\columnwidth]{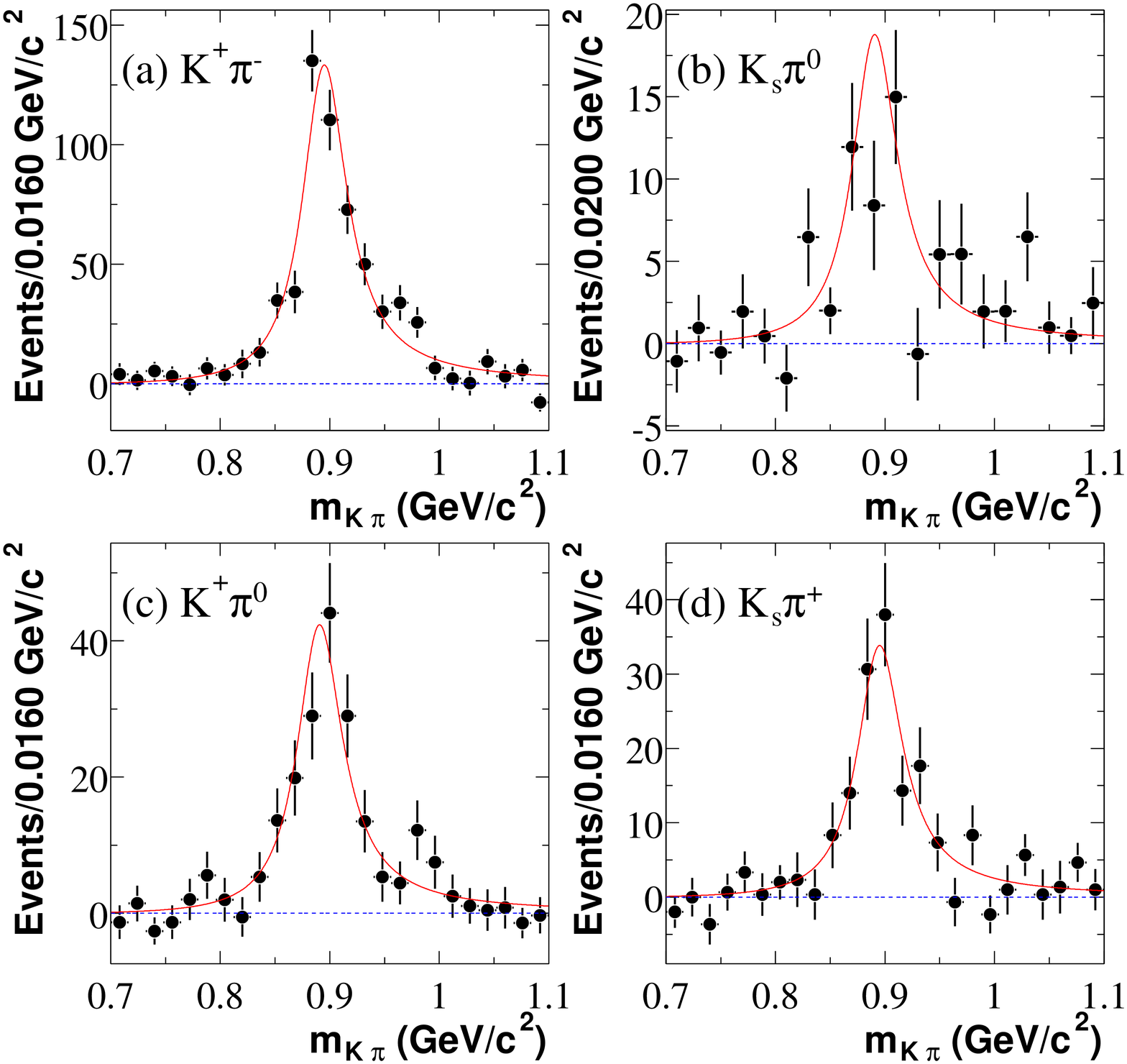}
\caption{$m_{K\pi}$ spectra for the different decay modes for events
in the signal region after background subtraction using sidebands in
$\mes$ and $\de$.  The points are data and solid curves represent
relativistic p-wave Breit-Wigner line shapes with masses and widths of
$K^*$ taken from Ref. ~\cite{PDBook}.}
\label{fig:kstar}
\end{figure}

\begin{table*}[ntb]
\begin{center}
\caption{The signal efficiency $\epsilon$, the fitted signal yield
$N_S$, the branching fraction ${\cal B}$, and the CP-asymmetry $\acp$
for each decay mode.  The combined branching fractions and
CP-asymmetries for $\bkgneut$ and for $\bkpg$ are also shown.  Errors
are statistical and systematic, with the exception of $\epsilon$ and
$N_S$, which have only systematic and statistical errors respectively.
The detailed systematic errors are listed in
Table~\ref{table:syst:eff}. }
\begin{tabular*}{\linewidth}{
@{\extracolsep{\fill}}l
@{\extracolsep{\fill}}c
@{\extracolsep{\fill}}c
@{\extracolsep{\fill}}c
@{\extracolsep{\fill}}c
@{\extracolsep{\fill}}c
@{\extracolsep{\fill}}c
}\hline\hline

    Mode&
    {$\epsilon$(\%)}&
    {$N_S$}&
    {${\cal B}(\times\mbox{10}^{-\mbox{5}}) $}&
    {Combined ${\cal B} (\times\mbox{10}^{-\mbox{5}}) $}&
    {${\acp}$}&
    {Combined ${\acp}$}\\\hline

    $K^+\pi^-$&
    24.4$\pm$1.4&
    583$\pm$30&
    3.92$\pm$0.20$\pm$0.23&
    \multirow{2}*{$\Bigr{\}}$3.92$\pm$0.20$\pm$0.24}&
    $-0.069 \pm 0.046 \pm 0.011$&
    \multirow{4}*{$\Biggr{\}}-0.013 \pm 0.036 \pm 0.010$}\\

    $K_s\pi^0$&
    15.3$\pm$1.9&
    \ 62$\pm$15&
    4.02$\pm$0.99$\pm$0.51&&&\\

    $K^+\pi^0$&
    17.4$\pm$1.6&
    251$\pm$23&
    4.90$\pm$0.45$\pm$0.46&
    \multirow{2}*{$\Bigr{\}}$3.87$\pm$0.28$\pm$0.26}&
    0.084$\pm$0.075$\pm$0.007&\\

    $K_s\pi^+$&
    22.1$\pm$1.4&
    157$\pm$16&
    3.52$\pm$0.35$\pm$0.22&
    &
    0.061$\pm$0.092$\pm$0.007&\\

    \hline\hline

\end{tabular*}
\end{center}
\label{table:results}
\end{table*}

\begin{table}
\caption{Fractional systematic uncertainties on the branching fractions
	 $\mathcal{B}$ and absolute systematic uncertainties on CP asymmetry $\acp$.}

        \begin{center}
        \begin{tabular*}{\columnwidth}{@{\extracolsep{\fill}}lrrrr}\hline\hline

    &
    \multicolumn{4}{c}{Systematic errors on $\mathcal{B}$(\%)}\\
     Description&
     {$K^+\pi^-$}&
     {$K_s^0\pi^0$}&
     {$K^+\pi^0$}&
     {$K^0_s\pi^+$}\\\hline

    Number of $B$ events        &1.1&1.1&1.1&1.1\\
    $R^{+/0}$                  &2.4&2.4&2.4&2.4\\
    Tracking efficiency         &1.6&   &0.8&0.8\\
    Charged particle identification         &1.0&   &1.0&1.0\\
    Photon efficiency           &2.5&7.6&7.6&2.5\\
    Photon isolation cut      &2.0&2.0&2.0&2.0\\
    $\pi^0, \eta\ $ veto    &1.0&1.0&1.0&1.0\\
    $K_s$ efficiency            &   &3.0&   &3.0\\
    Neural network          &3.0&3.5&2.7&2.8\\
    PDF parameterization        &2.2&7.3&2.7&1.4\\
    MC statistics/fit bias   	&0.9&3.2&2.4&1.6\\\hline
    Total                       &5.8&12.3&9.4&6.3\\\hline\hline

    &
    \multicolumn{4}{c}{Systematic errors on $\acp$ (\%)}\\

    Tracking efficiency             & 0.35  &   & 0.25  & 0.25 \\
    Charged particle identification & 1.00  &   & 0.55  & 0.53 \\ 
    Nuclear interaction asymmetry   & 0.20  &   & 0.35  & 0.15 \\ 
    $B$-background asymmetry        & 0.25  &   & 0.25  & 0.25 \\ \hline
    Total                           & 1.1   &   & 0.7   & 0.7 \\\hline\hline

    \end{tabular*}
    \end{center}
\label{table:syst:eff}
\end{table}

Background events arise predominantly from random combinations of
particles in \qqbar\ production ($q$$=$$u$,$d$,$s$,$c$), with the
high-energy photon originating from initial-state radiation or from
$\piz$ and $\eta$ decays.  We suppress this jet-like background in
favor of the spherical signal events, using several event-shape
variables as in Ref. \cite{Aubert:2001me}.  To maximize separation
between signal and background, these variables are combined in neural
networks that are separately optimized for each decay mode.  Each
network is trained using Monte Carlo (\MC) events, and is validated on
statistically independent \MC\ samples. Cuts are made on the
neural-network output to suppress continuum background.  The $\mes$
and $\de$ distributions of data are shown in Fig.~\ref{fig:mes} for
all four $\kstar$ decay modes.

The remaining background includes that from \BB\ events, which is
dominated by $B \to X_{s} \gamma$ decays, where $X_{s}$ represents
hadronic final states other than $K^*$.  If one or more particles
escape detection, $X_s$ may be incorrectly reconstructed as $K^*$,
leading to a value of $\mes$ near the B meson mass, but with
$\DeltaE^*$ distinctly negative.

For each decay mode, the signal yield and asymmetry $\acp$ (except for
the $K_s\pi^0$ mode) are simultaneously extracted using an extended
unbinned maximum likelihood fit,
$$
{\cal L}=\exp{\Bigl{(}-\sum_{i = 1}^{3} n_{i}\Bigr{)}}\cdot \Bigl{[}\prod_{j = 1}^{N}
\sum_{i=1}^3 N_i{\cal P}(\vec{x}_j;\ \vec{\alpha}_i)\Bigr{]},
$$
\noindent 
to the two-dimensional distribution of $m_{ES}$ and $\Delta E^*$ with
three hypotheses (index $i$): signal, continuum background, and $B$
background.  The probability density function (PDF) ${\cal
P}(\vec{x}_j;\ \vec{\alpha}_i)$ for each of the three hypotheses is
the product of individual PDFs of the fit variables
$\vec{x}_j=(m_{ES}, \Delta E^*)$.  $\vec{\alpha}_i$ are the shape
parameters for the PDFs described below.  In the three
self-flavor-tagged modes ($K^+\pi^-$, $K^+\pi^0$, and $K_s\pi^+$),
$N_i\ =\ \frac{1}{2}(1-{\cal F}{\acp}_i)n_i$, where $n_i$ and
${\acp}_i$ stand for the total yield and CP-asymmetry of signal,
continuum background, and $B$ background, while in the $K_s\pi^0$
decay mode, $N_i\ =\ n_i$.  The bottom-quark flavor, ${\cal F}$, is
defined as $-1$ for $b$ quarks and $+1$ for $\overline{b}$ quarks.  In
the $K^+\pi^-$ mode, mistagging is possible if both the pion and kaon
are misidentified, but this probability is negligibly small.  We
assume that the CP asymmetry of the $B$ background and that of the
continuum background are the same.

To reduce systematic errors, most of the fit parameters for the signal
and for the continuum background are determined by a fit to data.  For
continuum background, the $\DeltaE^*$ distribution is modeled by a
first-order polynomial function with the exception of $K_s\pi^+$,
where a second-order polynomial is used.  The $m_{ES}$ distribution
for continuum background is modeled with an ARGUS function
\cite{Albrecht:1990cs}.  In the $K^+\pi^0$ decay mode, the continuum
background shape is simultaneously fit to the off-resonance data to
obtain a stabler fit.  For the $B$ background, the Gaussian
distribution used for $\DeltaE^*$ and the Novosibirsk function
\cite{Novosibirsk} used for $m_{ES}$ have all shape parameters fixed
to values determined from \MC.  The signal $\DeltaE^*$ distribution is
modeled as a Crystal Ball function \cite{Bloom:1983}, which is a
Gaussian distribution with a low-side power-law tail that is fixed
using \MC.  The $m_{ES}$ distribution for signal is modeled as a
Gaussian function, except for the $K^+\pi^0$ decay mode, where a
Crystal Ball function, with tail parameters fixed using \MC\ fits, is
used to accommodate a low-side tail due to the $\pi^0$ energy lost
from the calorimeter.  The same low-side tail in the $K_s\pi^0$ decay
mode is ignored due to the small number of events in this mode.

Correlations between $\mes$ and $\de$ distributions could introduce a
bias in the signal yields. To study this, randomly selected events
from our detailed MC simulation of the signal were mixed with
background events generated using the PDF from the fit.  In this way
we determined that the $K^+ \pi^-$ efficiency must be corrected by
multiplying it by $0.98$. For the $K^0_S \pi^0$, $K^+\pi^0$, and
$K^0_S \pi^+$ modes, the corresponding numbers are $0.91$, $0.96$, and
$0.96$.  The error in this fit bias due to MC statistics is included
as a systematic uncertainty.  These MC studies also indicate that
correlations between the $B$ background and the continuum background
fit yields do not affect the fitted signal yield.

The projections of the maximum likelihood fits on $m_{ES}$ and $\Delta
E^*$ are shown in Fig.~\ref{fig:mes} for each decay mode.
Figure~\ref{fig:kstar} shows that the background-subtracted $K\pi$
invariant mass distributions agree well with the expected $K^*$
resonance shape.  This confirms that the signal is consistent with
coming from only true $K^*$ decays.

Table~\ref{table:results} shows signal efficiencies, yields from the
fits, and branching fractions (${\cal B}$) calculated using our recent
measurement \cite{Aubert:rpluszero} of the production ratio of charged
and neutral $B$ events, $R^{+/0}\equiv \Gamma(e^+e^-\rightarrow
B^+B^-)/ \Gamma(e^+e^-\rightarrow B^0\bar{B^0}) = 1.006 \pm 0.048$ at
$\sqrt{s}=M_{\Upsilon(4S)}$.

Combined values of $\cal{B}$(\bkog) and $\cal{B}$(\bkpg), which are
also shown in Table \ref{table:results}, are calculated taking into
account correlated systematic errors between modes.  We further
combined these measurements, using the lifetime ratio
$\tau_{B^+}/\tau_{B^0} = 1.083 \pm 0.017$ \cite{PDBook} and our
measurement of $R^{+/0}$, to find the isospin asymmetry, \isoval,
which corresponds to an allowed region of \isoexcl\ at the 90\%
confidence level.  We also present a combined $\acp$ measurement in
Table~\ref{table:results}, which corresponds to an allowed region of
\aexcl\ at the 90\% confidence level.

The systematic error on the branching fraction for each mode is shown
in Table~\ref{table:syst:eff}.  Most of the uncertainties are
determined as in our previous analysis ~\cite{Aubert:2001me}, so we
provide details only for the new procedures used.  The neural-network
inputs are generally independent of the fully reconstructed $\bkg$
candidate, so we determine their efficiencies and systematic
uncertainties with high-purity control samples with reconstructed
$\bpdpi$ and $\bzdpi$.  The ``PDF parameterization'' error comes from
\MC\ studies of our fitting procedure, in which we estimate the
uncertainty incurred by fixing parameters in the continuum and $B$
background models.  This includes uncertainty in the inclusive
branching fraction and spectral shape of $\incbsg$.

The systematic uncertainties in the measurement of $\acp$ are also
shown in Table~\ref{table:syst:eff}.  The first three contributions
arise from potential particle-antiparticle asymmetries in the detector
response, including differences in interaction cross-sections for
$K^+$ and $K^-$, and for $\pi^+$ and $\pi^-$ (estimated with a method
similar to that used in Ref.~\cite{Aubert:jpsik}).  The uncertainty
due to a possible asymmetry in the $B$ background, which is dominated
by $\incbsg$, is estimated by varying our recent measurement of
$\acp$($\incbsg$) \cite{Aubert:2004acp} within its errors.

We conclude that both the isospin- and CP-asymmetries in \bkg\ decay
processes are consistent with SM predictions.  The branching fractions
measured are also consistent with SM-based calculations and are more
precise than those predictions.  These measurements are consistent
with previous results \cite{Coan:1999kh,Aubert:2001me,Nakao:2004}.

We are grateful for the excellent luminosity and machine conditions
provided by our \pep2\ colleagues, and for the substantial dedicated
effort from the computing organizations that support \babar.  The
collaborating institutions wish to thank SLAC for its support and kind
hospitality.  This work is supported by DOE and NSF (USA), NSERC
(Canada), IHEP (China), CEA and CNRS-IN2P3 (France), BMBF and DFG
(Germany), INFN (Italy), FOM (The Netherlands), NFR (Norway), MIST
(Russia), and PPARC (United Kingdom).  Individuals have received
support from CONACyT (Mexico), A.~P.~Sloan Foundation, Research
Corporation, and Alexander von Humboldt Foundation.  

\bibliographystyle{h-physrev3}

\begin{thebibliography}{10}
\bibitem{Novosibirsk}
The Novosibirsk function is defined as $f(m_{ES}) = A_S \exp{( -0.5 \{ \ln^2[ 1
  + \Lambda\tau\cdot(m_{ES} - m_0) ]/\tau^2 +\tau^2 \} )}$, where $\Lambda =
  \sinh{( \tau \sqrt{\ln 4})}/{(\sigma \tau \sqrt{\ln 4})}$, the peak position
  is $m_0$, the width is $\sigma$, and $\tau$ is the tail parameter.
\end{thebibliography}

\end{document}